\documentclass[sigplan,screen,nonacm]{acmart}
\AtBeginDocument{%
  \providecommand\BibTeX{{%
    \normalfont B\kern-0.5em{\scshape i\kern-0.25em b}\kern-0.8em\TeX}}}

\usepackage{xspace}

\setcopyright{acmcopyright}
\copyrightyear{2018}
\acmYear{2018}
\acmDOI{XXXXXXX.XXXXXXX}

\acmConference[Conference acronym 'XX]{Make sure to enter the correct
  conference title from your rights confirmation email}{June 03--05,
  2018}{Woodstock, NY}
\acmPrice{15.00}
\acmISBN{978-1-4503-XXXX-X/18/06}

\newcommand{\longname}{\textsc{GitHub Recent Bugs}\xspace}
\newcommand{\name}{\textsc{GHRB}\xspace}

\begin{document}

\title{The GitHub Recent Bugs Dataset for Evaluating LLM-based Debugging Applications}

\author{Jae Yong Lee}
\affiliation{%
  \institution{KAIST}
  \city{Daejeon}
  \country{South Korea}
}

\author{Sungmin Kang}
\affiliation{%
  \institution{KAIST}
  \city{Daejeon}
  \country{South Korea}
}

\author{Juyeon Yoon}
\affiliation{%
  \institution{KAIST}
  \city{Daejeon}
  \country{South Korea}
}

\author{Shin Yoo}
\affiliation{%
  \institution{KAIST}
  \city{Daejeon}
  \country{South Korea}
}

\renewcommand{\shortauthors}{Lee et al.}

\begin{abstract}
Large Language Models (LLMs) have demonstrated strong natural language 
processing and code synthesis capabilities, which has led to their rapid 
adoption in software engineering applications. However, details about LLM 
training data are often not made public, which has caused concern as to whether 
existing bug benchmarks are included. In lieu of the training data for the 
popular GPT models, we examine the training data of the open-source LLM 
StarCoder, and find it likely that data from the widely used Defects4J 
benchmark was included, raising the possibility of its inclusion in GPT 
training data as well. This makes it difficult to tell how well LLM-based 
results on Defects4J would generalize, as for any results it would be unclear 
whether a technique's performance is due to LLM generalization or memorization. 
To remedy this issue and facilitate continued research on LLM-based SE, we 
present the GitHub Recent Bugs (GHRB) dataset, which includes 76 real-world 
Java bugs that were gathered after the OpenAI data cut-off point.
\end{abstract}


\maketitle

\section{Introduction}
\label{sec:introduction}

A significant portion of software engineering research revolves around the automatic 
detection~\cite{Manes2021FuzzingSurvey}, reproduction~\cite{Kang2023aa,Soltani2020EvoCrash}, and removal of 
software bugs~\cite{Gazzola2019APRSurvey}. Prior work has shown differences between artificial bugs and real-world 
bugs~\cite{Gopinath2014MutationFault,Just2014MutantSubstitute}, making real-world bugs valuable for research.
As a result,
software bug benchmarks using actual bugs from open-source software have been proposed, as such benchmarks allow standardized and fair evaluations of various techniques that deal with bugs.
Examples of such benchmarks include the 
widely used Defects4J~\cite{Just2014Defects4J}, Siemens~\cite{Do2005Siemens}, 
BugsInPy~\cite{Widyasari2020BugsInPy} and BugsJS~\cite{Gyimesi2019BugsJS} benchmarks. 

However, the permissive licenses that led to these benchmarks of real-world bugs also make these repositories a prime target as training data for code-based large language models. Large 
Language Models (LLMs), which are Transformer~\cite{Vaswani2017Attention} models with a large number of 
parameters, have been showing substantial performance gains relative to traditional techniques in multiple 
software engineering domains, such as in program repair~\cite{jiang2023impact,Fan2023LLMAPR} or test 
generation~\cite{Lemieux2023codamosa,Kang2023aa}. Having a large number of parameters, LLMs need a similarly 
large training dataset~\cite{Hoffmann2022ScalingChinchilla}, and thus
a large amount of software data from open-source repositories is gathered~\cite{Li2023StarCoder}. Indeed, 
our own findings show that bugs from the Defects4J benchmark are often included in the training data 
of the open-source LLM StarCoder~\cite{Li2023StarCoder}. While information on the training data for the most 
popular LLMs in use, such as ChatGPT from OpenAI, is not public~\cite{openai2023gpt4}, it is reasonable to 
assume that they would use similar training data.

The potential overlap between existing bug benchmarks and LLM training data raises a critical question: are the state-of-the-art results 
from LLMs due to the strengths of LLM generalization, or simply due to code memorization? While such a
concern had been voiced in early literature~\cite{chen2021evaluating}, we are 
unaware of subsequent attempts to build a real-world bug dataset that is not likely to be part of LLM 
training data.

To this end, we propose the \longname{} (\name{}) benchmark, which consists of 76 real-world Java bugs that
have been fixed after September 2021, which is the cut-off date of training data for OpenAI 
LLMs~\cite{OpenAIDataCutoff}, such as GPT-3.5 and GPT-4. We further confirm that these bugs were \emph{not} used for the training of the open-source StarCoder LLM. As a result, researchers can 
evaluate LLM-based applications on the \name benchmark without concern about data leakage.

In summary, our contributions are:

\begin{itemize}
\item We provide a partial evaluation into how pervasive data leakage may be for the widely-used real-world bug benchmark Defects4J, and report that there is likely significant data leakage when using LLMs.

\item We provide an automated framework to gather real-world bug benchmarks, and filter them so that only the most relevant are left for manual inspection.

\item We make our bug benchmark, \name, publicly available\footnote{\url{https://github.com/coinse/GHRB}}, which facilitates evaluation of LLM-based techniques that involve bugs without data leakage.
\end{itemize}

\begin{figure}
    \centering
    \includegraphics[width=0.9\linewidth]{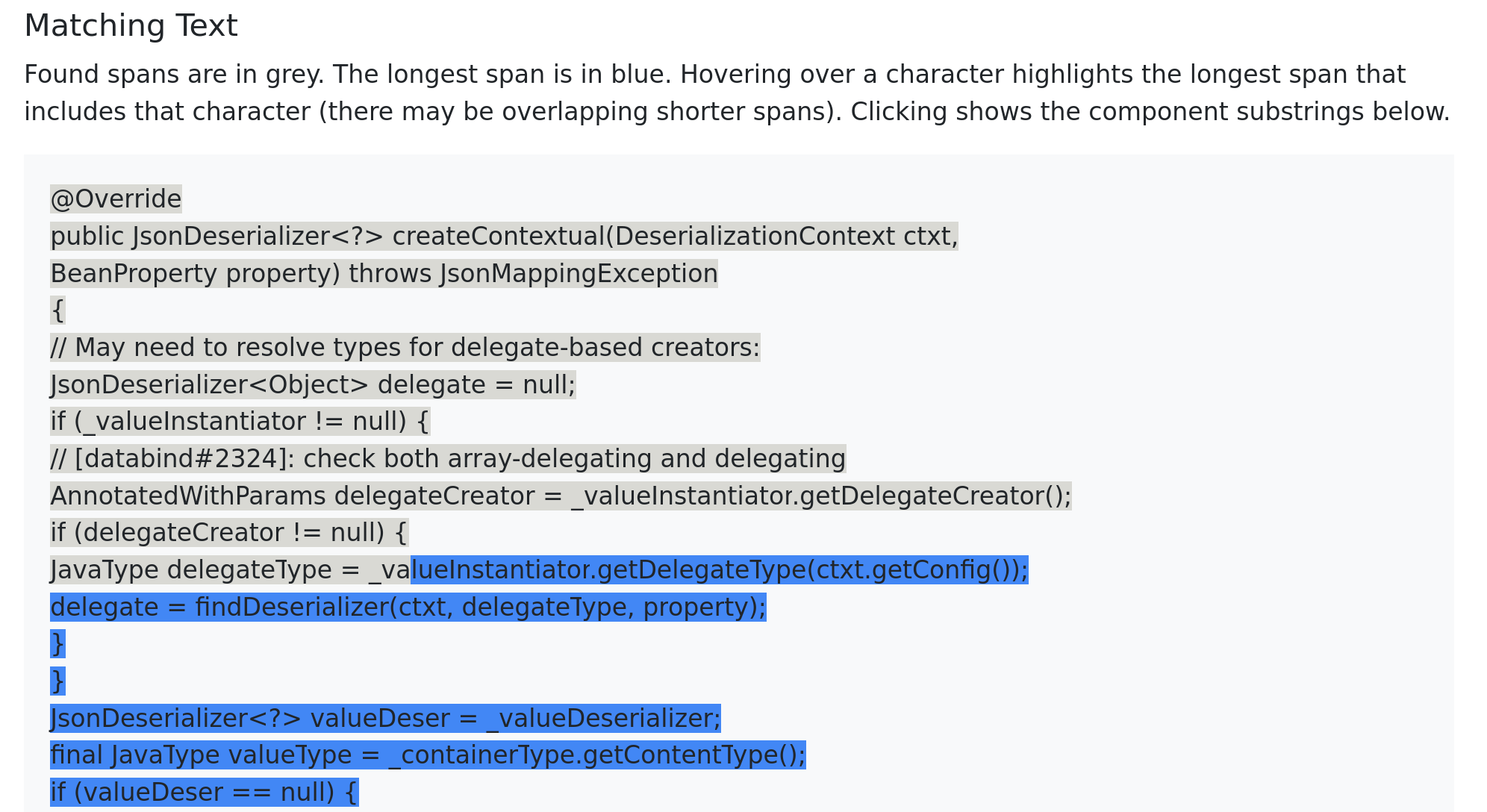} % JacksonDatabind-112
    \caption{A screenshot of the data inclusion website of StarCoder. Entire portions of the method are included, and the overlap break (gray to blue) happens at the fix location.}
    \label{fig:inclusion_screenshot}
\end{figure}

\section{Motivation}
\label{sec:motivation}

There has been a steady shift towards evaluating software engineering research artifacts using real-world bugs~\cite{Liu2019TBar,jiang2023impact,Koyuncu2019IfixR,Chen2018SequenceRSL,Sohn2017Fluccs}. These benchmarks allow comparisons of techniques on the same basis~\cite{Liu2020ot}. 
Further, evaluations based on these benchmarks, while not perfect, are likely to reveal the real-world 
performance of a given technique better than those based on artificially seeded faults.

\begin{table}[t!]
\caption{Compromised bugs by Defects4J v1.0 project}
\label{tab:compromise}
\scalebox{0.9}{
\begin{tabular}{lrrlrr}
\toprule
Project & \# Bugs & \% Comp. & Project & \# Bugs & \% Comp. \\
\midrule
Math    & 104 & 67.3\% & Lang    & 61  & 78.7\% \\
Time    & 25  & 92.0\% & Chart   & 25  & 72.0\% \\
Closure & 130 & 35.4\% &         &     &        \\
\bottomrule
\end{tabular}
}
\end{table}

The use of the same benchmarks has continued while LLM-based SE techniques 
have been rapidly introduced. Given the opacity of training details of OpenAI LLMs, it is difficult for 
researchers to assess whether LLMs have been trained on benchmark data. Instead, under the assumption that 
LLM developers are likely to gather similar training data, we provide an initial assessment of the degree to 
which existing benchmark data is included in open-source LLM training data. To do 
so, we use the `Data Portraits' tool\footnote{\url{https://stack.dataportraits.org/}}, 
which allows users to easily check whether any text was included in the StarCoder 
training data by highlighting the longest common subsequence of the input data that was also present in 
the training data.
Using this tool, we evaluate the widely-used Defects4J v1.0 benchmark. 
Specifically, we evaluate whether the bug-revealing test and the buggy method, two common artifacts used in 
LLM-based SE applications~\cite{Kang2023aa,Koyuncu2019IfixR,jiang2023impact}, are included in the training 
data; we consequently evaluate the 345 bugs with buggy methods. 
We conservatively assume that if 90\% of a test or method is included in StarCoder data, data about that
bug was likely included, making it inappropriate for evaluation when using StarCoder.

Worryingly, we find 35\% of tests and 39\% of buggy methods were included in StarCoder training data, using the 90\% criterion mentioned above. In combination, 59\% of Defects4J bugs were likely compromised. A breakdown of evaluation results by project is presented in Table~\ref{tab:compromise}.
Furthermore, some matched subsequences would `break' around 
the actual fix location, as in Figure~\ref{fig:inclusion_screenshot}, suggesting that 
StarCoder was trained with the fixed version of these methods. The new projects introduced in
Defects4J v2.0 were not safe either: the most recent bugs from all newly added projects 2.0 were also included following the same 
90\% criterion.
While it is difficult to know for sure, our finding strongly suggests that OpenAI may have included these projects in its LLM training 
data in a similar fashion, raising concerns about the validity of LLM-based SE evaluation using Defects4J.

Researchers have been aware of this issue~\cite{Kang2023aa,jiang2023impact,Fan2023LLMAPR}, with many developing new evaluation datasets. 
However, up to now 
these efforts have been ad-hoc and uncoordinated. We believe it would benefit LLM-based SE research if there 
existed a standard, real-world benchmark that was relatively free from data contamination concerns. 

\section{Data Collection}
\label{sec:process}

This section describes the process used to collect bugs for \longname (\name), that 
are recent enough to avoid being included in LLM training data. 
Specifically, every bug should meet the following requirements:

\noindent\textbf{1. The bug is in the source code:} Every bug in the database should exist inside the source 
code of the project, where the buggy and fixed version is explicitly marked by the contributors. 
Specifically, every bug should be related to the core functionalities of the project, and hence any fixes 
regarding the build scripts, build configurations, markdown documentations, and tests were deliberately 
excluded. 
    
\noindent\textbf{2. The bug is reproducible:} Every bug in the database should have at least one test that 
fails on the buggy version and passes on the fixed version. In other words, every bug should be accompanied 
by a bug revealing test.

\noindent\textbf{3. The bug is isolated:} For every bug in the database, the difference 
between the buggy and fixed versions should be directly related to the bug, and should not include any 
external changes such as feature additions or refactoring.

The scripts used to collect and verify our data are available in our artifact as well, facilitating expansion of \name.

\subsection{Identifying Potential Bugs}
\label{subsec:identifying}

A list of repositories was first compiled by combining the repositories in the initial GHRB dataset from Kang et al.~\cite{Kang2023aa} with the list of the 100 Java repositories with most stars from GitHub. For each repository, pull requests that were (1) created after the officially stated data cutoff point of 
OpenAI LLM models (September 2021), (2) reference a related bug report, and (3) either add or modify test 
files to introduce bug reproducing tests are automatically collected. Nonetheless, some pull requests
gathered as a result of this process were not bug fixes or described the bug in non-English languages.
To ensure the quality of our data, we filtered out pull requests that did not change Java files, used
the LangID~\cite{LangIDRepo} tool to ensure that the bug report was in English, and finally performed a manual inspection to check for consistency.

\subsection{Reproducing Bugs}
\label{subsec:reproducing}

All of the bugs in the database were filtered so that for each bug, we could identify at least one bug revealing test that fails on the buggy version but passes on the fixed version. Tests that (1) pass on both versions, (2) fail on both versions, or are (3) unrelated to the bug were removed during the process, via an automated process as well as a manual revision. Among the remaining tests, those that were ``flaky'', or those that exhibited non-deterministic behavior, were removed. Finally, the authors went through a manual revision to further filter out such bugs so that all the bugs included in the final \name dataset are reproducible to the best of our knowledge.

\section{Database of Real Bugs}
\label{sec:database}

The \name benchmark consists of 76 bugs from 16 repositories. The chosen repositories vary in size and popularity, 
but all are primarily written in Java. 
Table~\ref{tab:bugs} shows summary statistics of the dataset.

Similarly to our check in Section~\ref{sec:motivation}, we checked whether the \textit{oldest} bug-revealing tests gathered as part of \name were included in the StarCoder training data. We found no overlap over our 90\% criterion\footnote{The maximum overlap observed was 66\%: note that non-zero overlap is inevitable as these projects were created before September 2021.} used in Section~\ref{sec:motivation}, 
demonstrating that the bugs that \name contains are free from data contamination concerns when evaluating StarCoder-based applications. Since \name is based on commits merged after September 2021, \name is also `safe' 
when used to evaluate OpenAI LLM-based applications as well.

The \name dataset additionally provides the following:

\noindent\textbf{Metadata.} The bug database provides the creation date of the pull requests, the original bug report ID (ID of the pull request), and the URL of the bug report. 

\noindent\textbf{Bug revealing tests.} The bug database includes of the list of one or more tests that reveal the bug, which fails on the buggy version and passes on the fixed version. For each test, the absolute path and the root cause is available.

\noindent\textbf{Patch information.} The bug database includes of the patch information, which is collected by taking the {\itshape git diff} between the buggy and fixed version.

\section{Interface of \name}
\label{sec:abstraction}
In addition to the collection of bugs described in the previous section, the artifact of \name provides the following interfaces to facilitate ease of use.

\noindent\textbf{Interface to Version Control Systems:} The interface allow users to access the buggy and fixed versions of included repositories by using a simple flag, without the need of knowledge of the version control system adopted by the contributors. Each bug in the database is mapped to an integer ID, which is in the chronological order of its creation date. For each bug, the buggy and fixed version are denoted by the flags `b' for {\itshape buggy} and `f' for {\itshape fixed}, respectively. This creates a layer of abstraction over the version control system of each repository, as mentioned above.

\noindent\textbf{Interface to Build Environments:} The interface allows users to compile each target without the need to specify project-specific build environments. During runtime, \name automatically derives the required build tool (i.e., maven, gradle), the information whether the repository makes use of a project-specific script for compilation, and the specific build tool versions required to compile the target repository. This automatic identification is of particular convenience, as repositories tend to be built in varying versions of build tools and JDK. Some of the repositories required comparatively complex build environments, and for such cases the compilation wrapper scripts integrated in the original repository were used to specify the versions of tools. Overall, this abstraction relieves users from the burden of manually searching for build environments.

\noindent\textbf{Interface to Testing:} The interface allows users to test each compiled target without the need to specify the build environments nor the project-specific CLI commands. The \name interface automatically creates a configuration file when a user checkouts to a specific version of a repository. During testing, the interface automatically searches for the configuration file to derive the failing test information, which allows for efficient bug reproduction via test execution.

\section{Related Work}
\label{sec:relwork}

This work expands the bug reproduction dataset introduced by Kang et al.~\cite{Kang2023aa}. 
In that work, the authors constructed a real-world dataset consisting of 31 real-world Java bugs 
for which bug-reproducing tests were included after the OpenAI training data cutoff point to 
mitigate data leakage concerns. The \name of this publication significantly expands on that data, 
consists of 76 bugs from 16 repositories up from 31 bugs from 6 repositories in prior work, 
and provides a command-line interface that facilitates the use of \name.

This work is heavily inspired by prior bug benchmarks, 
as mentioned in Section~\ref{sec:introduction}, of which Defects4J is the primary example. 
As we show in 
Section~\ref{sec:motivation} there is data leakage risk when using it to evaluate 
algorithms using state-of-the-art LLMs. 
In contrast to Defects4J, \name exclusively consists of data created after the data cutoff point of 
OpenAI's LLMs\cite{OpenAIDataCutoff}, allowing LLM-based algorithms to be evaluated without concern of data contamination.
Other benchmarks exist in other languages as well, and we hope to expand our dataset to 
include other languages in future work.

\section{Conclusion}
\label{sec:conclusion}

We introduce \longname (\name), a real-world Java bug dataset designed to mitigate data leakage concerns when evaluating LLM-based software engineering techniques. Our hope is to present a supplementary evaluation benchmark to the larger and established bug benchmarks, so that software engineering researchers can evaluate the potential generality of their LLM-based tools more conveniently.

\begin{table*}[t!]
  \caption{Projects and number of real bugs available in the public release of \name (as of 6 September 2023)}
  \label{tab:bugs}
  \scalebox{0.9}{
  \begin{tabular}[width=0.9\linewidth]{lrrrrr|lrrrrr}
    \toprule
    Project & Bugs & LoC & Test LoC & \# Tests & \# Stars & Project & Bugs & LoC & Test LoC & \# Tests & \# Stars\\
    \midrule
    fastjson                 & 1  & 43.6k  & 143.4k & >435  & 25.4k & jackson-dataformat-xml & 1  & 5.9k   & 9.4k   & >2    & 541    \\
    nacos                    & 5  & 215.9k & 8.7k   & 2.7k  & 27.4k & gson                      & 12 & 9.0k   & 19.6k  & 1.3k  & 22.4k  \\
    dubbo                     & 1  & 146.7k & 89.4k  & 3.3k  & 39.3k & sslcontext               & 6  & 3.7k   & 7.2k   & 497   & 406    \\
    rocketmq                  & 12 & 150.3k & 54.7k  & 1.7k  & 19.8k & jsoup                        & 4  & 14.3k  & 12.5k  & 1.1k  & 10.3k  \\
    assertj                  & 4  & 45.9k  & 161.4k & 11.8k & 2.4k  & openapi-generator   & 5 & 9.8k   & 37.6k  & 1.8k  & 17.5k  \\
    checkstyle            & 15 & 41.7k  & 238.2k & 4.5k  & 7.8k  & seata                      & 2  & 166.4k & 30.5k  & 1.0k  & 24.2k  \\
    jackson-core           & 3  & 30.7k  & 44.8k  & >100  & 2.2k  & retrofit                  & 1  & 3.9k   & 6.5k   & 329   & 42k    \\
    jackson-databind       & 3  & 73.1k  & 71.0k  & >28   & 3.3k  & Apktool                   & 1 & 10.7k   & 3.4k   & 202 & 17.6k   \\ \midrule
           &  &   &   &    &   & \textbf{Total}            & \textbf{76} & \textbf{972}   & \textbf{938k}   & \textbf{30.8k} & \textbf{263k}   \\
  \bottomrule
\end{tabular}}
\end{table*}

\bibliographystyle{ACM-Reference-Format-num}
\interlinepenalty=10000
\bibliography{sample-base}

\end{document}